\documentclass[onecolumn,usenatbib]{mnras}

\usepackage{newtxtext,newtxmath}
\usepackage[T1]{fontenc}

\DeclareRobustCommand{\VAN}[3]{#2}
\let\VANthebibliography\thebibliography
\def\thebibliography{\DeclareRobustCommand{\VAN}[3]{##3}\VANthebibliography}

\newcommand{\xte}{XTE~J1810$-$197}
\newcommand{\maspy}{$\rm mas~yr^{-1}$}
\newcommand{\kmps}{$\rm km~s^{-1}$}
\newcommand{\mjypb}{$\rm mJy~beam^{-1}$}
\newcommand{\psrpi}{\ensuremath{\mathrm{PSR}\pi}}
\newcommand{\mspsrpi}{\ensuremath{\mathrm{MSPSR}\pi}}
\newcommand{\rcsl}{reduced chi-square}
\newcommand{\rcs}{$\chi_{\nu}^{2}$}
\newcommand{\multilinecomment}[1]{}
\newcommand{\snra}{G11.0$-$0.0}

\usepackage{graphicx}	% Including figure files
\usepackage{amsmath}	% Advanced maths commands
\usepackage{amssymb}	% Extra maths symbols
\title[VLBA Astrometry of \xte]{A magnetar parallax}

\author[H.~Ding et al.]{
H. Ding,$^{1,2}$\thanks{E-mail: haoding@swin.edu.au}
A. T. Deller,$^{1,2}$
M. E. Lower,$^{1,3}$
C. Flynn,$^{1,2}$
S. Chatterjee,$^{4}$
W. Brisken,$^{5}$
\newauthor
N. Hurley-Walker,$^{6}$
F. Camilo,$^{7}$
J. Sarkissian,$^{8}$
V. Gupta$^{1,2}$
\\
$^{1}$Centre for Astrophysics and Supercomputing, Swinburne University of Technology\\
John St, Hawthorn, VIC 3122, Australia\\
$^{2}$ARC Centre of Excellence for Gravitational Wave Discovery (OzGrav)\\
$^{3}$CSIRO Astronomy and Space Science, Australia Telescope National Facility, Epping, NSW 1710, Australia\\
$^{4}$Cornell Center for Astrophysics and Planetary Science and Department of Astronomy, Cornell University, Ithaca, NY 14853, USA\\
$^{5}$National Radio Astronomy Observatory, P.O. Box O, Socorro NM 87801, USA\\
$^{6}$International Centre for Radio Astronomy Research, Curtin University, Bentley, WA 6102, Australia\\
$^{7}$South African Radio Astronomy Observatory, 2 Fir Street, Observatory 7925, South Africa\\
$^{8}$CSIRO Astronomy and Space Science, Parkes Observatory, PO Box 276, Parkes NSW 2870, Australia
}

\date{Accepted XXX. Received YYY; in original form ZZZ}

\pubyear{2020}

\begin{document}
\label{firstpage}
\pagerange{\pageref{firstpage}--\pageref{lastpage}}
\maketitle

% Abstract of the paper
\begin{abstract}
\xte\ was the first magnetar identified to emit radio pulses, and has been extensively studied during a radio-bright phase in 2003--2008. 
It is estimated to be relatively nearby compared to other Galactic magnetars, and provides a useful prototype for the physics of high magnetic fields, magnetar velocities, and the plausible connection to extragalactic fast radio bursts.
Upon the re-brightening of the magnetar at radio wavelengths in late 2018, we resumed an astrometric campaign on \xte\ with the \textit{Very Long Baseline Array}, and sampled 14 new positions of \xte\ over 1.3 years. 
The phase calibration for the new observations was performed with two phase calibrators that are quasi-colinear  on the sky with \xte, enabling substantial improvement of the resultant astrometric precision.
Combining our new observations with two archival observations from 2006, we have refined the proper motion and reference position of the magnetar and have measured its annual geometric parallax, the first such measurement for a  magnetar. The parallax of $0.40\pm0.05$~mas corresponds to a most probable distance $2.5^{\,+0.4}_{\,-0.3}$\,kpc for \xte.
Our new astrometric results confirm an unremarkable transverse peculiar velocity of $\approx200$\,\kmps\ for \xte, which is only at the average level among the pulsar population. The magnetar proper motion vector points back to the central region of a supernova remnant (SNR) at a compatible distance at $\approx70$\,kyr ago, but a direct association is disfavored by the estimated SNR age of $\sim3$\,kyr.
\end{abstract}

% Select between one and six entries from the list of approved keywords.
% Don't make up new ones.

%%%%%%%%%%%%%%%%%%%%%%%%%%%%%%%%%%%%%%%%%%%%%%%%%%

%%%%%%%%%%%%%%%%% BODY OF PAPER %%%%%%%%%%%%%%%%%%

\section{Introduction}

Magnetars are a class of highly magnetized, slowly rotating neutron stars (NSs) with surface magnetic field strengths typically inferred in the range $10^{14}$--$10^{15}$\,G, making them the most magnetic objects in the known universe.
They have been observed to emit high energy electromagnetic radiation, and to undergo powerful X-ray and gamma-ray outbursts. The high energy emission from these objects is thought to be powered by the decay of their magnetic fields \citep{Thompson95} as opposed to dipole radiation for classical pulsars. 
To date, 29 magnetars and 6 magnetar candidates have been discovered \citep{Olausen14}\footnote{Catalogue: \url{http://www.physics.mcgill.ca/~pulsar/magnetar/main.html}}; however, only 6 magnetars have ever been observed to emit radio pulsations, partly due to a small birthrate for the class \citep{Gill07}.
SGR~J1935$+$2154 recently joined the other 5 magnetars that have been observed to emit radio pulses. Its radio emission was detected in the form of an unprecedented radio pulse with a fluence of $1.5\pm0.3$\,MJy~ms \citep[][]{Andersen20,Bochenek20}. That burst is the highest-fluence radio pulse ever recorded from the Galaxy and confirms magnetars are plausible sources of extragalactic Fast Radio Bursts (FRBs). 
However, the mechanism by which such strong radio pulses are produced from magnetars is poorly understood \citep[e.g.][]{Margalit20}, as is the birth mechanism of magnetars. 

Multi-wavelength observations of Galactic magnetars, including long-term timing and \textit{Neutron Star Interior Composition Explorer} (NICER) observations, allow us to study the morphology and evolution of their magnetic fields, and potentially probe their internal structure \citep{Kaspi17}.
However, such studies are usually limited by the uncertainties in the underlying magnetar distances (and uncertain proper motions as well, in some cases).
For instance, the X-ray spectrum fitting technique that has recently been made possible by observations with NICER requires a well-constrained, pre-determined distance to the target (which can be a magnetar) in order to infer its radius along with its mass \citep{Bogdanov19}. 
Besides, owing to the enormous instability in spin-down rates (period derivative $\dot{P}$) of magnetars \citep[e.g.][]{Camilo07,Archibald15,Scholz17}, measuring the proper motion (not to mention parallax) via timing is difficult for magnetars; as such, using an accurate, {\em a priori} proper motion and parallax in the timing analysis of a magnetar can improve the reliability of the timing model, thus facilitating the study of long-term $\dot{P}$ evolution.
Furthermore, an accurate distance would enable unbiased estimation of the absolute flux of X-ray flares or the absolute fluence of giant radio pulses. 
On top of the studies focusing on the magnetars, accurate distance and proper motion for a magnetar also enables constraints to be placed on the distance to the dominant foreground (scattering) interstellar-medium (ISM) screen (\citealp{Putney06}; see \citealp{Bower14,Bower15} for an example).

Proper motion measurements for magnetars are significant in their own right.
Both the space velocities of neutron stars and their surface magnetic field strengths have been connected to the progenitor stellar masses and the processes of core-collapse supernovae. 
\citet{Duncan92} suggested that the high magnetic fields of magnetars could be associated with very high space velocities of $\rm 10^3\,km~s^{-1}$ through e.g., asymmetric mass loss during core collapse or in the form of an anisotropic magnetized wind, or through a neutrino and/or photon rocket effect. Such processes would be ineffective for ``ordinary'' neutron star field strengths ($\lesssim 10^{13}$~G), leading to great interest in magnetar velocity estimates as diagnostics of their natal processes.
So far the transverse velocity measurements that have been made (modulo large uncertainties) do not support a higher-than-average kick velocity for magnetars, with transverse velocities around the range of 200\,\kmps\ inferred for \xte, PSR~J1550$-$5418, and PSR~J1745$-$2900 \citep[][respectively]{Helfand07,Deller12,Bower15}.
Additionally, the proper motion of a magnetar could provide a crucial test of its association with nearby supernova remnants (SNR), especially when the magnetar is outside the SNR. More importantly, the proper motion enables us to infer the kinematic age of the magnetar as well as the SNR from the underlying association, which is more reliable than the characteristic age of either.

Both distance and proper motion for a magnetar can be geometrically measured with VLBI (very long baseline interferometry) astrometry at radio wavelengths.
Observations of magnetar radio pulses reveal that they are quite distinct from the radio emission seen in pulsars -- most of them have flat radio spectra and their pulse profiles are highly variable on timescales ranging from seconds to years. Radio emission from magnetars is also a relatively short-lived phenomenon, generally starting out bright after an outburst, then fading over the following months to years. After radio emission ceases, they then spend long periods of time in a radio-silent, quiescent state before the next outburst.
With a current sample size of only 3 magnetars with precise VLBI proper motions, the (re-)appearance of a radio-emitting magnetar offers a valuable opportunity, particularly for pulsar timing and astrometry, both of which can be performed much more precisely with radio observations than with X-rays.

\xte\ (hereafter J1810) was discovered in 2003 due to an outburst at X-ray wavelengths \citep{Ibrahim04}, and was subsequently seen to be pulsating at radio wavelengths with a period of 5.4 seconds \citep{Camilo06}, the first time radio pulsations had been detected for a magnetar. During its brief period of radio brightness over a decade ago, two VLBA observations separated by 106 days were made, allowing the measurement of a proper motion of $13.5 \pm 1.0$\,\maspy\ -- also the first for a magnetar \citep{Helfand07}. The implied transverse velocity was $212 \pm 35$\,km\,s$^{-1}$ (assuming a distance of $3.5 \pm 0.5$\,kpc) and was the first indication that magnetar velocities might be much lower than the expectations outlined earlier in this section.
J1810 subsequently faded into a 10 year quiescence at radio wavelengths until 2018 December 8, when it was found to be radio bright (and pulsating) again at Jodrell Bank \citep{Lyne18}, Molonglo Radio Observatory \citep{Lower18} and Effelsberg \citep{Desvignes18}. The flux density of the source ranged between 9 and 20~mJy from 835\,MHz to 8\,GHz, showing a flat spectrum. 

Here we present new VLBI astrometric results for J1810 in Section~\ref{sec:results}, and lay out the direct indications of the results in Section~\ref{sec:discussion}. We also describe dual-calibrator phase calibration technique (also known as 1-D interpolation \citealp{Fomalont03}) and detail the relevant data analysis in Sections~\ref{sec:data_reduction} and \ref{sec:results}.
Throughout this paper, the uncertainties are provided at 68\% confidence level unless otherwise stated.

\section{Observations and correlation}
\label{sec:observations}

After the re-activation of J1810 at radio wavelengths in December 2018, we observed the magnetar with the \textit{Very Long Baseline Array} (VLBA) from January 2019 to November 2019 on a monthly basis (project code BD223).
We re-visited J1810 with three consecutive VLBA observations in the same observing setup on 28 March, 6 April and 13 April 2020 (when J1810 was at its parallax maximum; project code BD231). Altogether there are 14 new VLBA observation epochs.

\begin{figure}
    \centering
	\includegraphics[width=10cm]{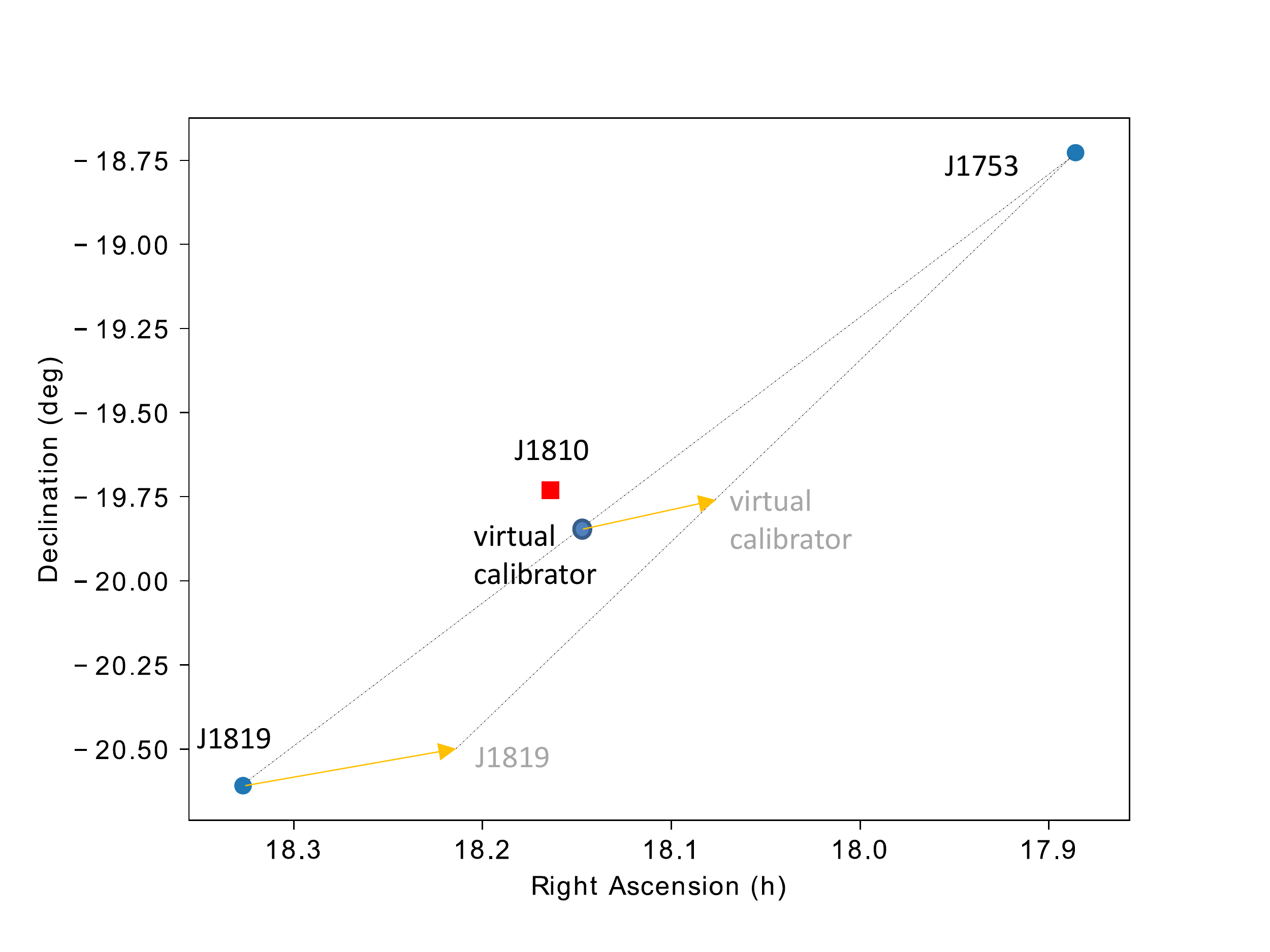}
    \caption{
    {\bf a)} The diagram shows the positions of the magnetar (marked with red rectangle), the primary phase calibrator J1753 and secondary phase calibrator J1819. The virtual calibrator is chosen as the closest point (12\farcm5 away) to J1810 on the J1753-to-J1819 geodesic line, which is at 62.43\% of the distance to J1819 from J1753. {\bf b)} Yellow arrows are overlaid on the calibrator plan to illustrate the discussion in Sections~\ref{subsec:systematics} and \ref{subsec:absolute_position}: if the position for J1819 was in error (greatly exaggerated here for visual effect -- any position offset is in reality only at the $\sim1$\,mas level), the position of the virtual calibrator would change in the same direction.  The degree of the position shift of the virtual calibrator would be 62.4\% that of J1819. Any position error for J1753 would likewise affect the position of the virtual calibrator.
    }
    \label{fig:calibrator_plan}
\end{figure}

All the observations were carried out at around 5.7~GHz in astrometric phase referencing mode (where the pointing of the array alternates between the target and phase calibrator throughout the observation). Unlike typical astrometric observations, J1810 was phase-referenced to two phase calibrators: ICRF~J175309.0-184338 (4\fdg1 away from J1810, hereafter J1753) and VCS6~J1819-2036 (2\fdg5 away from J1810, hereafter J1819), which are almost colinear with J1810 (see Figure~\ref{fig:calibrator_plan}). The purpose of this non-standard astrometric setup is explained in Section~\ref{subsec:dualphscal}.
ICRF~J173302.7$-$130449 was observed as the fringe finder.

The data were correlated with the {\tt DiFX} software correlator \citep{Deller11a} in two passes, standard (ungated) and gated. For the gated pass, the off-pulse durations were excluded from the correlation in order to improve the signal-to-noise ratio (S/N) on the magnetar. For all the 14 observations, pulsar gating was applied based on the pulsar ephemerides obtained with our monitoring observations of the magnetar at the Parkes and Molonglo telescopes.
The timing observations at Parkes were carried out under the project code P885; the relevant observing setup and data reduction are described in Lower et~al. (in preparation).
The timing observations at Molonglo were fulfilled as part of the UTMOST project \citep{Jankowski19,Lower20}.

\begin{figure}
    \centering
	\includegraphics[width=12cm]{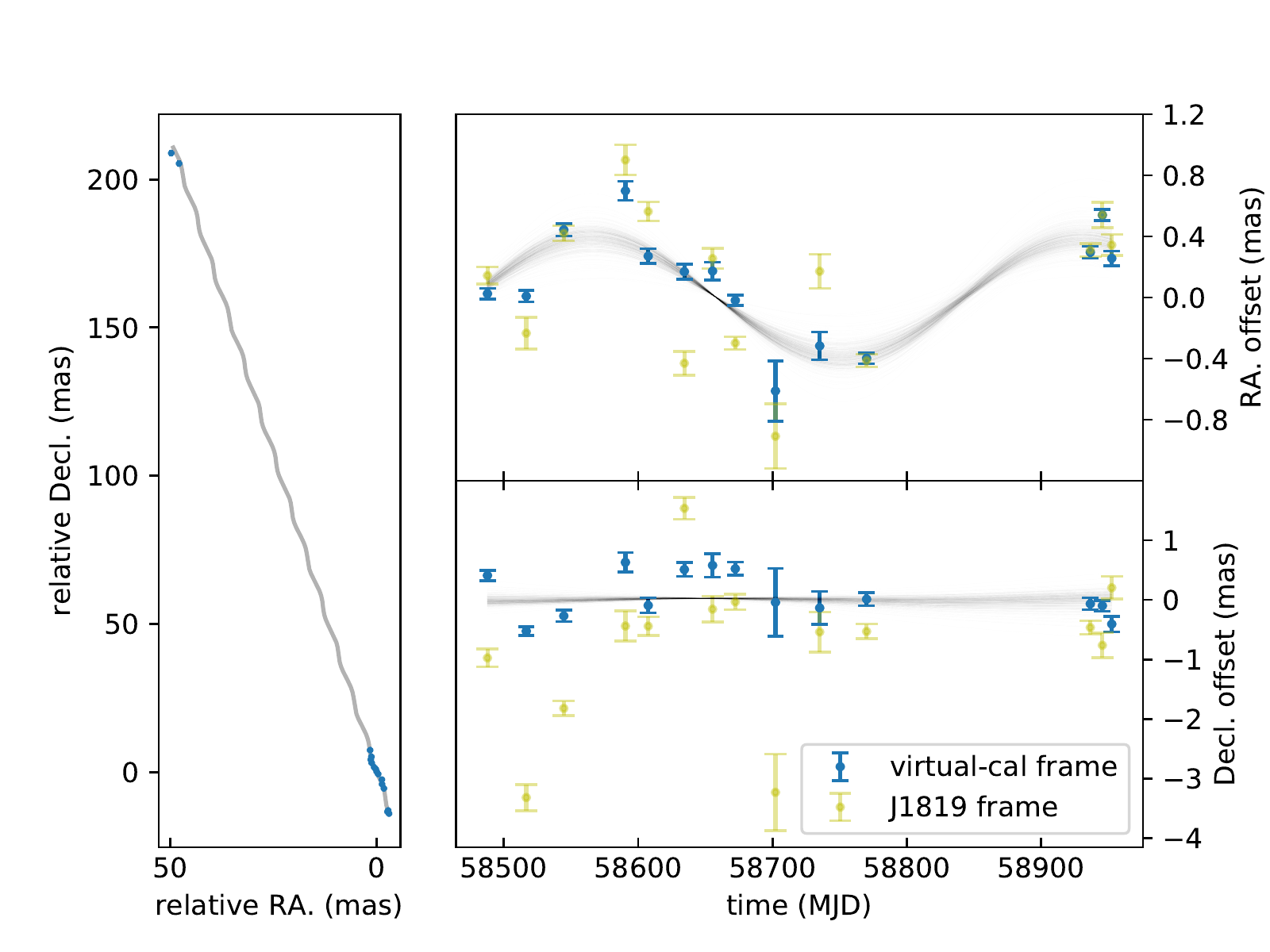}
    \caption{
    {\bf Left panel:} position evolution of J1810 relative to $18^{\rm h}09^{\rm m}51\fs 08333, -19\degr43'52\farcs1398$
    from 2006 to 2020. 
    {\bf Right panel:} recent positions of J1810 with proper motion subtracted measured in the virtual-calibrator frame (blue) and J1819 frame (yellow; see Section~\ref{subsec:reference_frames} for explanations of different reference frames); to convenience visual comparison, the two sets of positions adopt the same systematic uncertainties at each epoch as obtained in the virtual-calibrator frame.
    Overlaid are the best-fit models for 500 bootstrap draws from the positions measured in the virtual-calibrator frame.
    Here, the systematic uncertainties for the positions measured in the virtual-calibrator frame are still under-estimated, which is addressed in Section~\ref{subsec:pi_and_mu} using a bootstrap technique. 
    Obviously, the positions measured in the virtual-calibrator frame provide tighter constraints to the model.}
    \label{fig:positions_and_model}
\end{figure}

Apart from the 14 new VLBA observations, we re-visited two archival VLBA observations of J1810 taken in 2006 under the project code BH142 and BH145A \citep{Helfand07}.
An overview of observation dates is provided in Figure~\ref{fig:positions_and_model}. The two observations in 2006 were carried out at both 5\,GHz and 8.4\,GHz, using J1753 as the only phase calibrator. More details of the observing setup for the two observations in 2006 can be found in \citet{Helfand07}.
Hereafter, where unambiguous, positions obtained from the two epochs in 2006 and the 14 new epochs are equally referred to, respectively, as the ``year-2006 positions" and the ``recent positions".

\section{Data reduction}
\label{sec:data_reduction}
All VLBI data were reduced with the {\tt psrvlbireduce} (\url{https://github.com/dingswin/psrvlbireduce}) pipeline written in python-based {\tt ParselTongue} \citep{Kettenis06}. {\tt ParselTongue} serves as an interface to interact with {\tt AIPS} \citep{Greisen03} and {\tt DIFMAP} \citep{Shepherd94}. 
The pipeline has been incrementally developed for VLBI pulsar observations over the last decade, notably for the large VLBA programs \psrpi\ \citep{Deller19} and \mspsrpi\ \citep[e.g.][]{Ding20}.

For the two year-2006 epochs, only data taken at 5\,GHz were reduced and analyzed, in order to avoid potential position uncertainties due to any  frequency-dependent core shift in the phase calibrators \citep[e.g.][]{Bartel86,Lobanov98}.
All positions of J1810 were obtained from the gated J1810 data, the S/N of which exceed the ungated J1810 data by 40\% on average.

Due to multi-path scattering caused by the turbulent ISM along the line of sight to J1810 or J1819, the deconvolved angular size of J1810 is mildly broadened by $0.7\pm0.4$\,mas (detailed in Appendix~\ref{sec:scatter_broadening}), and J1810 (as well as J1819) is fainter at longer baselines. In addition, J1819 exhibits intrinsic source structure, with a jet-like feature extending over $\sim$10~mas.  Accordingly, J1819 is heavily resolved by the longest baselines of the VLBA, with a flux density of at most a couple of mJy on baselines longer than 50~M$\lambda$ (mega-wavelength).
The three stations furthest from the geographic centre of VLBA are MK (Mauna Kea), SC (St. Croix) and HN (Hancock), each of which only has 0--1 baselines shorter than 50~M$\lambda$.  Unsurprisingly, we found that while valid solutions could still often be found, the phase solutions for HN, MK, and SC are much noisier than for the remainder of the array.  
On top of this, atmospheric fluctuation causes larger phase variation at longer baselines, which, combined with the worse phase solutions, makes phase wraps (to be explained in Section~\ref{subsec:dualphscal}) at MK, SC and HN hard to determine.
We found that the image S/N for J1810 generally improves when MK, SC and HN are excluded from the target field data, and as a result, taking the three stations out of the J1810 data leads to a statistical positional error comparable to that obtained when using the full array.
Therefore, we consistently flagged the three stations from the final J1810 data for 14 recent epochs. However, we did not remove the three stations from any calibration steps, because we found the participation of the three stations allows better performance of self-calibration on J1819 at other stations.

The applied calibration steps in this work are largely the same as the \psrpi\ project \citep{Deller19} except for the phase calibration.
For the two year-2006 epochs, the phase solutions obtained with J1753 were directly transferred to J1810; whereas, for the recent 14 epochs, phase solutions were corrected based on the positions of J1753 and J1819, before being applied to J1810. Such a technique, though previously adopted by other researchers \citep[e.g.][]{Fomalont03}, is applied to a pulsar for the first time.

\subsection{Dual-calibrator phase calibration}
\label{subsec:dualphscal}
During the phase calibration of VLBI data, the calibration step $k~(k=2, 3, ...)$ provides an increment of the phase difference $\Delta\phi^{(k)}_{n}(t)$ between the station $n$ and the reference station in the VLBI array at a given time $t$ (for simplicity, frequency-dependency is not accounted for), which is then added to the previous sum of the phase difference $\phi^{(k-1)}_{n}(t)$ when the new solutions are applied.
The phase calibration of the primary phase calibrator (J1753 for this work) is followed by the self-calibration of the secondary phase calibrator (J1819 for this work), the phase solutions of which are predominantly limited by anisotropic atmospheric (including ionosphere and troposphere) propagation effect. 
Accordingly, the position-dependent phase difference should be formulated as $\Delta\phi_{n}(\vec{x},t)$, where $\vec{x}$ represents the 2-D sky position. 
In the normal phase calibration, the solutions $\Delta\phi_{n}(\vec{x}_\mathcal{S},t)$ obtained with the self-calibration of the secondary phase calibrator at $\vec{x}_\mathcal{S}$ are directly given to the target. The closer $\vec{x}_\mathcal{S}$ is to the target field, the better $\Delta\phi_{n}(\vec{x}_\mathcal{S},t)$ can approximate the unknown $\Delta\phi_{n}(\vec{x}_\mathcal{T},t)$ at the target field. 
In cases where $\vec{x}_\mathcal{S}$ is $\gtrsim1$\degr\ away from the target field, considerable offsets are expected between $\Delta\phi_{n}(\vec{x}_\mathcal{S},t)$ and $\Delta\phi_{n}(\vec{x}_\mathcal{T},t)$ (especially at lower frequencies), which are commonly treated as systematic errors when using the normal phase calibration. 
However, if the primary phase calibrator, secondary phase calibrator and the target happen to be quasi-colinear on the sky, $\Delta\phi_{n}(\vec{x}_\mathcal{T},t)$ can be well approximated by phase solutions corrected from $\Delta\phi_{n}(\vec{x}_\mathcal{S},t)$ \citep{Fomalont03}.

It is easy to prove that for three arbitrary different co-linear positions $\vec{x}$, $\vec{x}_1$ and $\vec{x}_2$,
\begin{equation}
    \phi_{n}(\vec{x},t)=\frac{\vec{x}_1-\vec{x}}{\vec{x}_1-\vec{x}_2}\phi_{n}(\vec{x}_2,t)+\frac{\vec{x}-\vec{x}_2}{\vec{x}_1-\vec{x}_2}\phi_{n}(\vec{x}_1,t),
	\label{eq:taylor_expansion}
\end{equation}
assuming higher-than-first-order terms are negligible (as supported by \citealp{Chatterjee04,Kirsten15}). Specific to the self-calibration of the secondary phase calibrator, we have $\Delta\phi_{n}(\vec{x},t)=\Delta\phi_{n}(\vec{x}_\mathcal{S},t)\cdot (\vec{x}_\mathcal{P}-\vec{x})/(\vec{x}_\mathcal{P}-\vec{x}_\mathcal{S})$, where $\vec{x}_\mathcal{P}$ and $\vec{x}_\mathcal{S}$ refer to the position of the primary phase calibrator and secondary phase calibrator, respectively; this relation allows us to extrapolate to $\Delta\phi_{n}(\vec{x},t)$ at any position $\vec{x}$ colinear with J1753 and J1819 based on $\Delta\phi_{n}(\vec{x}_\mathrm{J1819},t)$. 

We calculated the closest position to J1810 on the J1753-to-J1819 geodesic line (an arc), where it can best approximate $\Delta\phi_{n}(\vec{x}_\mathrm{J1810},t)$.
Since it acts like a phase calibrator for J1810, hereafter we term it the ``virtual calibrator'' for J1810.
The position of the virtual calibrator is $\vec{x}_\mathrm{v}=0.38\cdot \vec{x}_\mathrm{J1753}+0.62\cdot \vec{x}_\mathrm{J1819}$, which is 12\farcm5 away from J1810 (see Figure~\ref{fig:calibrator_plan}). Accordingly, the phase solutions extrapolated to the virtual calibrator are $\Delta\phi_{n}(\vec{x}_\mathrm{v},t)=0.62\cdot \Delta\phi_{n}(\vec{x}_\mathrm{J1819},t)$;
using this relation, we corrected the phase solutions of J1819 obtained with its self-calibration. The correction was implemented with a dedicated module called {\tt calibrate\_target\_phase\_with\_two\_colinear\_phscals} that was newly added to the {\tt vlbatasks.py}, as a part of the {\tt psrvlbireduce} package. 
One of the functions of the module is to solve the phase ambiguity of $\Delta\phi_{n}(\vec{x}_\mathrm{J1819},t)$ prior to multiplying $\Delta\phi_{n}(\vec{x}_\mathrm{J1819},t)$ by 0.62.

The biggest challenge of the dual-calibrator phase calibration is the phase ambiguity of $\Delta\phi_{n}(\vec{x}_\mathrm{J1819},t)$, which can be equivalently expressed as $\Delta\phi_{n}(\vec{x}_\mathrm{J1819},t)\pm2i\pi~(i=0,1,2,...)$. This phase ambiguity will not change the quality of solutions for the normal phase calibration, but will cause trouble for the dual-calibrator phase calibration, as the periodicity of the phase is broken when multiplied by a factor. 
The degree of phase ambiguity depends on observing frequency and angular distance between the main and secondary phase calibrator;
for this work (5.7\,GHz, 6.5\degr), we run into mild phase ambiguity, mainly at the longest baselines (that we do not use anyway as mentioned earlier in this section).
Among the recent epochs, the smallest size of the synthesized beam excluding MK, SC and HN is $2.9\times8.2$\,mas, more than two times larger than the level of the systematic uncertainties dominated by propagation effect (see Figure~\ref{fig:positions_and_model}).
Therefore, for this work, we consider $\Delta\phi_{n}(\vec{x}_\mathrm{J1819},t)$ less likely to turn more than one wrap, and impossible to turn more than two wraps (i.e. $|i|\leq2$).

We resolved the phase ambiguity of $\Delta\phi_{n}(\vec{x}_\mathrm{J1819},t)$ in a semi-automatic and iterative manner. The pipeline would go though all values of $\Delta\phi_{n}(\vec{x}_\mathrm{J1819},t)$ at each station. If $|\Delta\phi_{n}(\vec{x}_\mathrm{J1819},t)|<\pi/2$ holds true throughout the observation, then the solutions $\Delta\phi_{n}(\vec{x}_\mathrm{J1819},t)$ are deemed phase-unambiguous, and no human intervention is needed for the station $n$. Otherwise, $\Delta\phi_{n}(\vec{x}_\mathrm{J1819},t)$ solutions are plotted out for inspection and interactive correction. 
In most cases, no interactive correction is necessary after the inspection of the $\Delta\phi_{n}(\vec{x}_\mathrm{J1819},t)$ plot, as the solutions look continuous, oscillating around 0 within a reasonable range (e.g. between $\pm2\pi/3$). 
In the few cases where interactive corrections are needed, solutions are ambiguous in phase for at most 1 or 2 stations per observation. 
This enabled us to take a simple, brute-force approach to trialing the plausible possibilities for phase wraps (adding or subtracting an integer multiple by $2\pi$ radians to the solution) with interactive correction. For every possibility, we implemented the dual-calibrator phase calibration using the resultant solution and ran through the complete (data-reduction and imaging) pipeline. The correct $\Delta\phi_{n}(\vec{x}_\mathrm{v},t)$ should outperform other possibilities in terms of the image S/N for J1810 (and the S/N difference is normally significant), as it better approximates the $\Delta\phi_{n}(\vec{x}_\mathrm{J1810},t)$. This S/N criterium helps us find the ``real'' $\Delta\phi_{n}(\vec{x}_\mathrm{J1819},t)$, thus the right $\Delta\phi_{n}(\vec{x}_\mathrm{v},t)$.
The obtained solutions $\Delta\phi_{n}(\vec{x}_\mathrm{v},t)$ were then transferred to J1810.

More generally, if no phase-calibrator pair quasi-colinear with the target is found, one can extrapolate to $\Delta\phi_{n}(\vec{x},t)$ at any position $\vec{x}$ with three non-colinear phase calibrators \citep[also known as 2-D interpolation,][]{Fomalont03,Rioja17}.
Despite the longer observing cycle and hence sparser time-domain sampling \citep[unless using the multi-view observing setup,][]{Rioja17}, the tri-calibrator phase calibration can in principle remove all the first-order position-dependent systematics. 

\section{Systematic Errors and Astrometric Fits}
\label{sec:results}

\subsection{Reference frames in relative VLBI astrometry}
\label{subsec:reference_frames}

Similar to the way a reference frame is normally defined in non-relativistic (Cartesian) contexts, a reference frame in the context of relative VLBI astrometry (hereafter reference frame or frame) generally refers to a system of an infinite amount of sky positions that are tied to a phase calibrator (not necessarily a real one), in which positions are measured relative to the phase calibrator.
In this work there are three different reference frames where we can measure the positions of J1810: the J1753 frame, the J1819 frame, and the virtual-calibrator frame.
To be more specific, in the J1753/J1819 frame, the positions are measured relative to the brightest spot of the model image for J1753/J1819 respectively. By applying an identical model of J1753/J1819\footnote{available at \url{https://data-portal.hpc.swin.edu.au/dataset/calibrator-models-used-for-vlba-astrometry-of-xte-j1810-197}} during the fringe fitting and self-calibration steps, the J1753/J1819 images at different epochs are aligned, respectively,
to $17^{\rm h}53^{\rm m}09\fs 0886, -18\degr43'38\farcs520$
and $18^{\rm h}19^{\rm m}36\fs 8955, -20\degr36'31\farcs573$;
the virtual-calibrator frame is thus anchored to $18^{\rm h}09^{\rm m}35\fs 9437, -19\degr55'49\farcs656$,
determined by the relation $\vec{x}_\mathrm{v}=0.38\cdot \vec{x}_\mathrm{J1753}+0.62\cdot \vec{x}_\mathrm{J1819}$.
For the 14 recent epochs, the final positions of J1810 were measured in the virtual-calibrator frame, though the positions of J1810 were also measured in the other two frames for various purposes (see Section~\ref{subsec:systematics} and Figure~\ref{fig:positions_and_model}).
The two year-2006 positions were merely measured in the J1753 frame, as J1753 is the only available phase calibrator for these observations.

\subsection{Systematic errors and frame transformation}
\label{subsec:systematics}
Similar to the way in which systematic positional errors for pulsars in the \psrpi\ project were evaluated using Eqn~1 of \citet{Deller19} to account for both differential ionospheric propagation effects and thermal noise at secondary phase calibrators,
the estimation of systematic uncertainty for the measured positions of J1810 is based on the mathematical formalism
\begin{equation}
\label{eq:empirical_sys_error}
{\Delta_\mathrm{sys}}^2=\left( A \cdot \frac{s}{1\,{\rm arcmin}} \cdot \overline{\csc{\epsilon}}\right)^2 + \left(B/S\right)^2\,,
\end{equation}
where $\Delta_\mathrm{sys}$ is the ratio of the systematic error to the synthesized beam size, $\epsilon$ stands for elevation angle, $\overline{\csc{\epsilon}}$ is the average $\csc{\epsilon}$ for a given observation (over time and antennas), $s$ is the angular separation between J1810 and the calibrator of the frame, $S$ represents the image S/N of the calibrator of the frame, and $A$ and $B$ are coefficients to be determined.
However, unlike Eqn~1 of \citet{Deller19}, in Eqn~\ref{eq:empirical_sys_error} the two contributing terms on the right side are added in quadrature.  Given that they should be uncorrelated, this is more appropriate than the linear summation in Eqn~1 of \citet{Deller19}.

For this work, the second term of Eqn~\ref{eq:empirical_sys_error} is negligible for any of the three frames, as both J1753 and J1819 are strong sources, with a brightness $\geq$24\,\mjypb\ at our typical resolution (after MK, SC and HN have been removed from the array). 
In order to find a reasonable estimate of $A$ for this work, we measured the positions of J1810 consistently in the J1753 frame for all 16 epochs, and determined the value of $A$ that renders an astrometric fit (see ``direct fitting'' in Section~\ref{subsec:pi_and_mu}) with unity \rcsl, or \rcs~$=1$. We obtained $A=3\times10^{-4}$.
We note that, in principle, $A$ is invariant with respect to different reference frames. 
As is mentioned in Section~\ref{subsec:reference_frames}, the final positions of J1810 were measured in the virtual-calibrator frame for the 14 recent epochs and in the J1753 frame for the two year-2006 epochs.
Using $A=3\times10^{-4}$ and Eqn~\ref{eq:empirical_sys_error} without the second term, we acquired systematic errors for each epoch, which was then added in quadrature to the random errors.

After the determination of the systematic errors, the next step is to transform positions into the same reference frame. 
Since J1753 and J1819 are remote quasars almost static in the sky, the frame transformation is simply translational.
We translated the two year-2006 positions measured from the J1753 to the virtual-calibrator frame. The translation is equivalent, but in the reverse direction, to translate from the virtual-calibrator frame to the J1753 frame, which is easier to comprehend. In order to translate the virtual-calibrator frame to the J1753 frame, the position of the virtual calibrator needs to be measured in the J1753 frame, which can be accomplished by measuring the position of J1819 in the J1753 frame. The method to estimate the position of J1819 $\vec{x}'_\mathrm{J1819}$ and its uncertainty $\vec{\sigma}'_\mathrm{J1819}$ in the J1753 frame is detailed in Section~3.2 of \citet{Ding20}.

As is shown in Figure~\ref{fig:calibrator_plan}, once $\vec{x}'_\mathrm{J1819}$ is measured in the J1753 frame, the new position of the virtual calibrator $\vec{x}'_\mathrm{v}$ in the J1753 frame is also determined, the uncertainty $\vec{\sigma}'_\mathrm{v}$ of which is 0.62 times $\vec{\sigma}'_\mathrm{J1819}$. The difference between $\vec{x}'_\mathrm{v}$ and $\vec{x}_\mathrm{v}$ (or 0.62 times the difference between $\vec{x}'_\mathrm{J1819}$ and $\vec{x}_\mathrm{J1819}$) was used to translate the two year-2006 positions of J1810 from the J1753 frame to the virtual-calibrator frame. The $\vec{\sigma}'_\mathrm{v}$ was added in quadrature to the error budget (already including systematic and random errors) of the two year-2006 positions.

\subsection{Proper motion, parallax and distance}
\label{subsec:pi_and_mu}
After including the systematic errors and unifying to the virtual-calibrator frame, the 16 positions of J1810 can be used for astrometric fitting. 
Astrometric fitting was performed using {\tt pmpar}\footnote{\url{https://github.com/walterfb/pmpar}}. The median among the 16 epochs, MJD~58645, was adopted as the reference epoch. The results out of direct fitting are reproduced in Table~\ref{tab:mu_and_pi}, the \rcs\ of which is 10.6. The large \rcs\ suggests the systematic errors for the recent 14 positions are probably under-estimated, and the actual uncertainty for either parallax or proper motion is about 3 times larger than the uncertainty from direct fitting. 

\begin{figure}
    \centering
	\includegraphics[width=17cm]{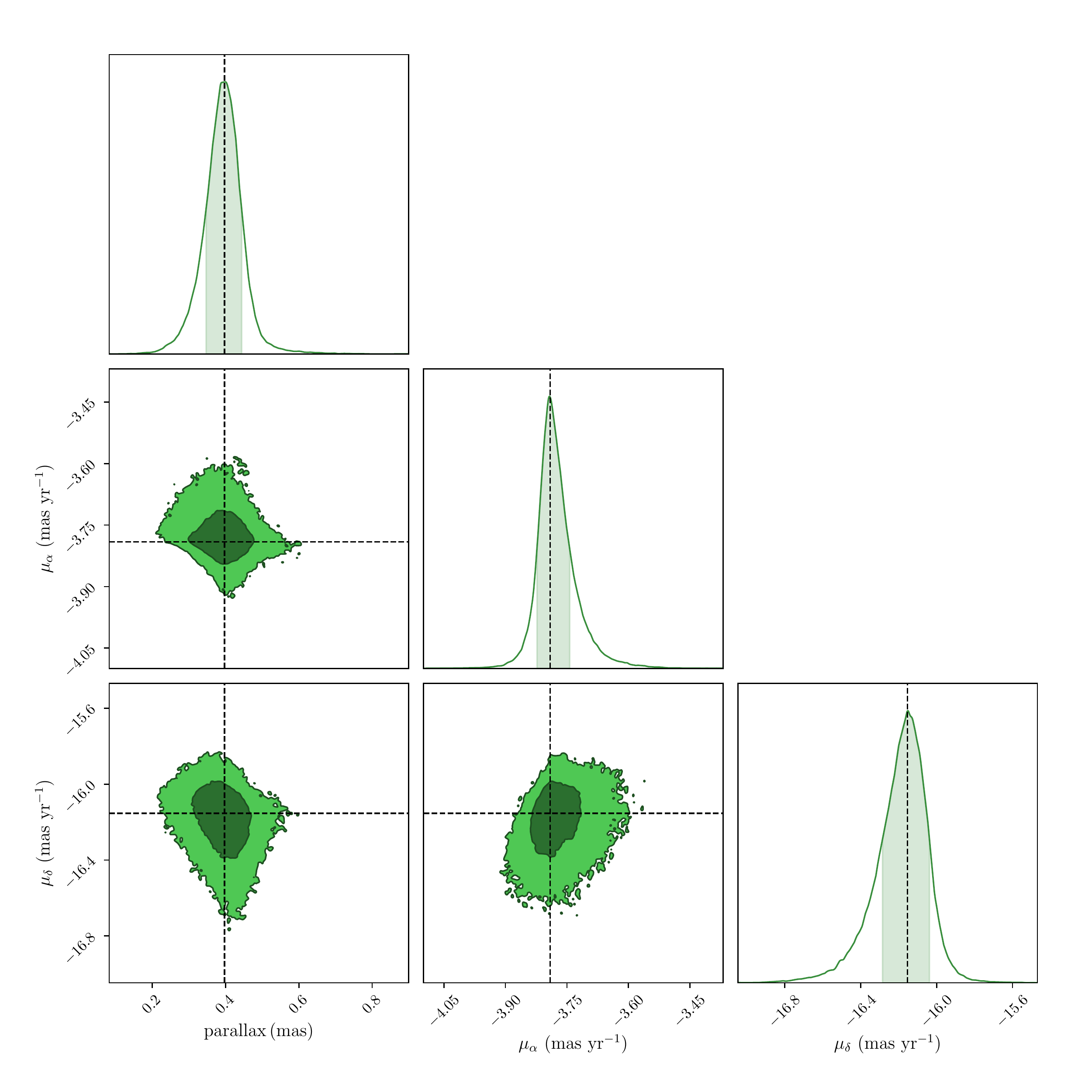}
    \caption{Error ``ellipses'' and marginalized histograms for parallax and proper motion.
    In each histogram, the dashed line marks the measured value; the shade stands for the 68\% confidence interval.
    In each error ``ellipse'', the dark and bright contour enclose, respectively, 68\% and 95\% of the bootstrapped data points.}
    \label{fig:covariance_pi_and_mu}
\end{figure}

Applying a bootstrap technique to astrometry can generally provide more conservative uncertainties, compared to direct fitting \citep[e.g.][]{Deller19}. 
In the same way as is described in Section~3.1 on \citet{Ding20}, we bootstrapped 100000 times, from which we assembled 100000 fitted parallaxes, proper motions and reference positions for J1810. The marginalized histograms for parallax and proper motion as well as their paired error ``ellipses'' are displayed in Figure~\ref{fig:covariance_pi_and_mu}. 
We reported the most probable value at the peak of each histogram as the measured value; the most compact interval containing 68\% of the sample was taken as the 68\% uncertainty range of the measured value (see Figure~\ref{fig:covariance_pi_and_mu}). The parallax and proper motion estimated with bootstrap are listed in Table~\ref{tab:mu_and_pi}, which are highly consistent with direct fitting while over 3 times more conservative (as is expected from the \rcs\ of direct fitting). Thus, the precision achieved for parallax and proper motion gauged with bootstrap can be deemed reasonable.
The parallax corresponds to the distance $2.5^{\,+0.4}_{\,-0.3}$\,kpc. Such a precision in distance would not be achieved with normal phase calibration using the same data (see Figure~\ref{fig:positions_and_model}).

\begin{table}
	\centering
	\caption{Proper motion and distance measurements for J1810}
	\label{tab:mu_and_pi}
	\begin{tabular}{cccccc} % four columns, alignment for each
		\hline
	method & $\mu_\alpha \equiv \dot{\alpha}\cos{\delta}$ & $\mu_\delta$ & $\varpi$ & $D$ & References \\
	    & (\maspy) & (\maspy) & (mas) & (kpc) & \\
		\hline
	direct fitting	& $-3.78\pm0.01$ & $-16.18\pm0.03$ & $0.39\pm0.01$ & $2.5\pm0.1$ & this work\\
	bootstrap	& $-3.79^{\,+0.05}_{\,-0.03}$ & $-16.2\pm0.1$ & $0.40\pm0.05$ & $2.5^{\,+0.4}_{\,-0.3}$  & this work\\
	\\
	Previous VLBI astrometry & $-6.60\pm0.06$ & $-11.72\pm1.03$ & $-$ & $-$ & \citet{Helfand07} \\
	red clump stars & $-$ & $-$ & $-$ & $3.1\pm0.5$ & \citet{Durant06} \\
	neutral hydrogen absorption & $-$ & $-$ & $-$ & 3.1$-$4.0 & \citet{Minter08} \\
		\hline
	\end{tabular}
\end{table}

\subsection{Absolute position}
\label{subsec:absolute_position}
Along with proper motion and parallax, a reference position $18^{\rm h}09^{\rm m}51\fs 083326 \pm 0.03\,{\rm mas}, -19\degr43'52\farcs1398 \pm 0.1$\,mas at the reference epoch MJD~58645 was also obtained for J1810 with bootstrap.
We note again the reference position was measured in the virtual-calibrator frame. According to the relation $\vec{x}_\mathrm{v}=0.38\cdot \vec{x}_\mathrm{J1753}+0.62\cdot \vec{x}_\mathrm{J1819}$, change in $\vec{x}_\mathrm{J1753}$ or $\vec{x}_\mathrm{J1819}$ would cause the position shift of the virtual calibrator $\Delta \vec{x}_\mathrm{v}$ (hence the position of J1810), following the relation 
\begin{equation}
\label{eq:position_shift_of_virtual_cal}
\Delta \vec{x}_\mathrm{v}=0.38\cdot \Delta \vec{x}_\mathrm{J1753}+0.62\cdot \Delta \vec{x}_\mathrm{J1819}\,.  
\end{equation}
Using Eqn~\ref{eq:position_shift_of_virtual_cal} and the method outlined in Section~3.2 on \citet{Ding20}, the reference position was shifted to align with the latest positions of J1753 and J1819\footnote{\url{http://astrogeo.org/vlbi/solutions/rfc_2020b/rfc_2020b_cat.html}}. 
The shifted reference position $18^{\rm h}09^{\rm m}51\fs 08333 \pm 0.3\,{\rm mas}, -19\degr43'52\farcs1418 \pm~0.5$\,mas is the absolute position of J1810 at MJD~58645, where the uncertainties of the J1753 and J1819 positions have been propagated onto the uncertainty budget.
At 5.7\,GHz, the effect of frequency-dependent core shift \citep[e.g.][]{Bartel86,Lobanov98} is at the $\lesssim0.1$\,mas level in each direction \citep{Sokolovsky11}, which makes unnoticeable difference to the uncertainty of the absolute position.

\section{Discussion}
\label{sec:discussion}
As is shown in Table~\ref{tab:mu_and_pi}, our new proper motion significantly improves on the previous value inferred from the two year-2006 positions;
the new distance $D=2.5^{\,+0.4}_{\,-0.3}$\,kpc is consistent with $3.1\pm0.5$\,kpc estimated using red clump stars \citep{Durant06}, while in mild tension with 3.1$-$4.0\,kpc constrained with neutral-hydrogen absorption \citep{Minter08}, suggesting the distance to the neutral-hydrogen screen was over-estimated.

In models of NS kicks from the electromagnetic rocket effect \citep{Harrison75} one might expect magnetars to have higher velocities \citep{Duncan92}.
Our new parallax and proper motion corresponds to the transverse velocity $v_t=198^{\,+29}_{\,-23}$\,\kmps. Using the Galactic geometric parameters provided by \citet{Reid19} and assuming a flat rotation curve between J1810 and the Sun, the peculiar velocity (with respect to the neighbourhood of J1810) perpendicular to the line of sight was calculated to be $v_{b}=-54\pm8$\,\kmps\ and $v_{l}=-175\pm26$\,\kmps. Our refined astrometric results consolidate the conclusion by \citet{Helfand07} that J1810 has a peculiar velocity typically seen in ``normal'' pulsars, unless its radial velocity is several times larger than the transverse velocity.

\subsection{SNR Association}
\label{subsec:snr_association}
The closest cataloged SNR to J1810 is \snra\ \citep{Green19}\footnote{\url{http://www.mrao.cam.ac.uk/surveys/snrs/}}, a partial-shell SNR $9'\times11'$ in size \citep{Brogan04,Brogan06}. The position of its geometric center is $18^{\rm h}10^{\rm m}04^{\rm s}, -19\degr25'$, $19'$ away from J1810. 
The latest distance estimate of \snra\ by \citet{Shan18} is $2.4\pm0.7$\,kpc, consistent with our new distance of J1810. Using our astrometric results, we find that the projected position at 70\,kyr ago is $18^{\rm h}10^{\rm m}09\fs 8, -19\degr25'00''$, about $1'$ east to the geometric center of \snra. For the above geometric reasons, it is possible that J1810 is associated with \snra. 

The plausibility of this potential association can be tested by considering the ages of both J1810 and \snra.
The spin-down rate $\dot{P}$ is erratic for J1810 \citep{Camilo07} as well as other magnetars \citep{Archibald15,Scholz17}, making the characteristic age $\tau_\mathrm{c}$ an unreliable estimate of the true age for J1810. Over the course of a decade, the changing value of $\dot{P}$  for J1810 has led to the $\tau_\mathrm{c}$ ($\propto 1/\dot{P}$) increasing from 11\,kyr \citep{Camilo07} to 31\,kyr \citep{Pintore18}. 
While the characteristic age is currently less than the tentative kinematic age $\tau^{*}_\mathrm{k}$ that the tentative association would imply, the unreliability of the $\tau_\mathrm{c}$ estimator in the case of magnetars suggests that the association cannot be ruled out on this basis.

From the perspective of \snra, the compactness of the SNR (see Figure~1 of \citealp{Castelletti16}) suggests that it is probably in the Sedov-Taylor stage. In this stage, the relation between the SNR radius $R_\mathrm{SNR}$ and its age $\tau_\mathrm{SNR}$ can be rewritten from \citet{Sedov59} as
\begin{equation}
R_\mathrm{SNR}\approx5\left(\frac{E}{10^{51}\,\mathrm{erg}}\right)^{1/5}\left(\frac{n}{30\,\mathrm{cm}^{-3}}\right)^{-1/5}\left(\frac{\tau_\mathrm{SNR}}{1\,\mathrm{kyr}}\right)^{2/5}\,\mathrm{pc}\,,
\label{eq:snr_r2}
\end{equation}
where the injected energy $E$ is expressed in a value typical of spherical SNRs expanding into the Galactic ISM, and the ambient ISM density $n\sim30\,\mathrm{cm}^{-3}$ for the $\gamma$-ray-emitting region including \snra\ was required to power the observed $\gamma$-ray emission above 1\,TeV at a distance of 2.4\,kpc \citep{Castelletti16}.
At an SNR distance $D_\mathrm{SNR}=2.4$\,kpc \citep{Shan18}, $R_\mathrm{SNR}\leq$3.8\,pc (corresponding to the angular size along the long axis), which yields $\tau_\mathrm{SNR}\lesssim3$\,kyr using Eqn~\ref{eq:snr_r2}, consistent with an SNR in the early part of the Sedov-Taylor phase. 
A $\tau_\mathrm{SNR}$ of 70\,kyr can be made possible with an injected energy 500 times smaller than the typically-assumed value of $10^{51}$\,erg, which is extremely unlikely \citep{Leahy17}. 
Therefore, we conclude \snra\ is not directly associated with J1810.
This is not too surprising, as less than half of the known magnetar population has a potential SNR association \citep{Olausen14}. Additionally, it has been proposed a strong post-birth magnetar wind can accelerate the dissipation of the SNR \citep{Duncan92}.

Though \snra\ is not directly associated with J1810, the 3-D geometric alignment might not be a coincidence. One explanation for the geometric alignment is: the progenitor star of J1810 has been in orbit with another supergiant companion;
and \snra\ is the SNR for the ``divorced'' companion of J1810.
In such a scenario, the progenitor of J1810 underwent a supernova explosion $\approx70$\,kyr ago and became unbound from its original companion.  The companion star continued evolving in isolation before itself undergoing a supernova explosion at $\lesssim3$\,kyr ago.
In this ``companion SNR'' scenario, assuming the components of the stellar system were formed at approximately the same time, the progenitor of J1810 should be slightly more massive than its companion (and hence evolve faster). 
However, given that this scenario would require that the companion underwent a supernova explosion only $\approx67$\,kyr (compared to the typical supergiant age of $\gtrsim1$\,Myr) after the first supernova, the mass difference of the two progenitor stars would have to be small. 

Assuming no peculiar velocity of the progenitor binary system (with regard to its neighbourhood mean) as well as a flat rotation curve between the Sun and J1810, the expected proper motion of the barycentre of the supergiant binary as observed from the Earth is only 1.2\,\maspy. The additional proper motion of the companion due to the orbital motion at the moment of unbinding is even smaller for supergiant binaries. Thus, the accumulated position shift of the companion star after the unbinding is at the $1.3'$ level across 67\,kyr, which does not violate the premise of the ``companion SNR'' scenario.

In principle, given the large age scatter of supergiants, the ``companion SNR'' scenario allows J1810 to be indirectly associated with an SNR further away. For example, J1810 can be traced back to 17' west to the centre of SNR~G11.4$-$0.1 (which is though far from the boundary of the SNR) at $\approx144$\,kyr ago. However, the relatively small characteristic age $\tau_\mathrm{c}=11$--31\,kyr favors the closer indirect association (or no association) with \snra.
Despite the ``companion SNR'' scenario, we note that it is highly possible that J1810 does not come from the \snra\ region; instead, it comes from an already dissipated SNR between J1810 and \snra\ (as supported by \citealp{Duncan92}).  
Longer-term $\tau_\mathrm{c}$ monitoring with timing observations on J1810 will offer a more credible range of $\tau_\mathrm{c}$ to be compared with the tentative kinematic age $\tau^{*}_\mathrm{k}\approx70$\,kyr suggested by the possible ``indirect'' association between \snra\ and J1810. Besides, a deeper search for SNR in the narrow region between J1810 and \snra\ might provide an alternative candidate for the SNR associated with J1810.

\section*{Acknowledgements}

We thank Marten van Kerkwijk for his in-depth review and helpful comments on this paper.
H.D. is supported by the ACAMAR (Australia-ChinA ConsortiuM for Astrophysical Research) scholarship, which is partly funded by the China Scholarship Council (CSC).
A.T.D is the recipient of an ARC Future Fellowship (FT150100415).
S.C. acknowledges support from the National Science Foundation (AAG~1815242).
Parts of this research were conducted by the Australian Research Council Centre of Excellence for Gravitational Wave Discovery (OzGrav), through project number CE170100004.
This work is based on observations with the Very Long Baseline Array (VLBA), which is operated by the National Radio Astronomy Observatory (NRAO). The NRAO is a facility of the National Science Foundation operated under cooperative agreement by Associated Universities, Inc.
Data reduction and analysis was performed on OzSTAR, the Swinburne-based supercomputer.
This work made use of the Swinburne University of Technology software correlator, developed as part of the Australian Major National Research Facilities Programme and operated under license.

\section*{Data and code availability}
\label{sec:data_availability}
The pipeline for data reduction is available at \url{https://github.com/dingswin/psrvlbireduce}.\\
All VLBA data used in this work can be found at \url{https://archive.nrao.edu/archive/advquery.jsp} under the project codes bd223, bd231, bh142 and bh145a.
The calibrator models for J1753 and J1819 can be downloaded from \\ \url{https://data-portal.hpc.swin.edu.au/dataset/calibrator-models-used-for-vlba-astrometry-of-xte-j1810-197}.

%%%%%%%%%%%%%%%%%%%%%%%%%%%%%%%%%%%%%%%%%%%%%%%%%%

%%%%%%%%%%%%%%%%%%%% REFERENCES %%%%%%%%%%%%%%%%%%

% The best way to enter references is to use BibTeX:

\bibliographystyle{mnras}
\bibliography{XTEJ1810-197,haoding} % if your bibtex file is called example.bib

% Alternatively you could enter them by hand, like this:
% This method is tedious and prone to error if you have lots of references
%\begin{thebibliography}{99}
%\bibitem[\protect\citeauthoryear{Author}{2012}]{Author2012}
%Author A.~N., 2013, Journal of Improbable Astronomy, 1, 1
%\bibitem[\protect\citeauthoryear{Others}{2013}]{Others2013}
%Others S., 2012, Journal of Interesting Stuff, 17, 198
%\end{thebibliography}

%%%%%%%%%%%%%%%%%%%%%%%%%%%%%%%%%%%%%%%%%%%%%%%%%%

%%%%%%%%%%%%%%%%% APPENDICES %%%%%%%%%%%%%%%%%%%%%

\appendix

\section{Measuring scatter-broadened size of XTE~J1810--197}
\label{sec:scatter_broadening}
The angular size of a radio source can be measured from its image deconvolved by the synthesized beam.
As magnetars are point-like radio sources, a non-zero deconvolved angular size of J1810 can be attributed to scatter-broadening effect caused by ISM.
For each epoch, the deconvolved image of J1810 is obtained as an elliptical gaussian component; the mean of its major- and minor-axis lengths is used as the scatter-broadened size of J1810. Assuming the degree of scatter-broadening did not vary across the 14 recent epochs, the 14 measurements of scatter-broadened sizes yield a scatter-broadened size of $0.7\pm0.4$\,mas for J1810.

%If you want to present additional material which would interrupt the flow of the main paper, it can be placed in an Appendix which appears after the list of references.

%%%%%%%%%%%%%%%%%%%%%%%%%%%%%%%%%%%%%%%%%%%%%%%%%%

% Don't change these lines
\bsp	% typesetting comment
\label{lastpage}
\end{document}